\documentclass[twocolumn, twocolappendix]{aastex7}

\usepackage{amsmath}
\usepackage{amssymb}
\usepackage{bm}
\usepackage{xcolor}

\received{August 15, 2025}
\revised{October 3, 2025}
\accepted{October 3, 2025}

\submitjournal{ApJ}

\newcommand{\parens}[1]{\left(#1\right)}
\newcommand{\brackets}[1]{\left[#1\right]}

\graphicspath{{./}{figs/}}
\begin{document}

\title{Advanced Weights for IXPE Polarization Analysis}

\author[0000-0002-6401-778X]{Jack T. Dinsmore}
\email{jtd@stanford.edu}
\affiliation{Department of Physics, Stanford University, Stanford CA 94305}
\affiliation{Kavli Institute for Particle Astrophysics and Cosmology, Stanford University, Stanford CA 94305}
\author[0000-0001-6711-3286]{Roger W. Romani}
\email{rwr@astro.stanford.edu}
\affiliation{Department of Physics, Stanford University, Stanford CA 94305}
\affiliation{Kavli Institute for Particle Astrophysics and Cosmology, Stanford University, Stanford CA 94305}

\begin{abstract}
As the Imaging X-ray Polarimetry Explorer (IXPE) measures increasingly faint sources, the need for precise polarimetry extraction becomes paramount. In addition to previously described neural-net (NN) weights, we introduce here point-spread function weights and particle background weights, which can be critical for faint sources. In some cases these can be augmented by time/phase and energy weights. We provide a publicly available analysis tool to incorporate these new weights, validate our method on simulated data, and test it on archival IXPE observations. Together these weights decrease the area of the polarization uncertainty contour by a factor of two compared to baseline IXPE analysis and will be essential for background-limited IXPE observations.
\end{abstract}

\section{Introduction} \label{sec:intro}
The Imaging X-ray Polarimetry Explorer \citep[IXPE,][]{weisskopf2022imaging, soffitta2021instrument} provides unique insights into the geometry of active galactic nuclei, X-ray binaries, pulsar wind nebulae (PWNe) and other X-ray sources. This advance comes from the track images which IXPE's gas pixel detectors \citep[GPDs,][]{baldini2021design} provide for each measured X-ray. Much has been achieved with the analytic mission-standard moments (Mom) track analysis which provides estimates of the events' position, energy, electric vector polarization angle (EVPA), and EVPA accuracy. As we observe fainter IXPE targets, exposures become very long and background contamination becomes highly significant; advanced methods to more optimally exploit the track image information become increasingly attractive.

For example \cite{peirson2021deep} and \cite{cibrario2025mitigating} introduce neural net (NN)-based methods to extract event parameters from IXPE tracks. Since photon and particle event morphologies differ, \cite{dimarco2023handling} suggest cuts based on track metadata that partly suppress particle background. Spectro-polarimetric fitting uses event energies to improve polarization analysis \citep[e.g.~\texttt{3ML},][]{vianello2015multi}, and IXPE's excellent timing precision may be used to separate a variable source from a constant background \citep[e.g.~the `simultaneous fitting' of][]{wong2023improved}. Measurements best constrain source polarization when maximum likelihood estimators \citep[MLE,][]{marshall2021multiband, gonzalez2023unbinned, marshall2024further} are applied. Though useful, many of these methods are not compatible, as presently formulated. Furthermore, additional refinements are possible. Prior X-ray observations generally provide detailed image morphology of IXPE targets. With accurate IXPE response functions, these measurements allow IXPE event positions and energies to be used as weights, concentrating the statistical power of IXPE data onto the polarization properties.

In this work, we develop a system of track-based weights to improve extraction of source polarization. The NN-technique of \citep{peirson2021deep} already provides effective polarization weights. We introduce a new NN which provides weights for particle rejection. We also use new, accurate on-orbit IXPE PSFs \citep[deconvolved from point source observations;][]{dinsmore2024polarization} to provide spatial weights. These weights correct for ``polarization leakage'' \citep{bucciantini2023polarization}, spurious polarized fringes introduced by correlations in errors between the reconstructed event position and EVPA \citep[as described by][]{dinsmore2024leakagelib}. \S\ref{sec:methods} details these and temporal/phase and energy weights, and presents their incorporation into a unified MLE analysis. \S\ref{sec:injection} uses simulations of a polarized point source to show large improvement over the mission-standard extraction, and insensitivity to systematics. We apply the method to published IXPE data in \S\ref{sec:data}, again demonstrating improved uncertainties, and conclude with a brief discussion in \S\ref{sec:conclusion}.

\section{Methods} \label{sec:methods}
The likelihood $L$---the probability distribution function (PDF) of the observed data given the model parameters---may be factored into a product of likelihoods for each event as follows:
\begin{equation}
    L = \prod_{\mathrm{event}\, i} \sum_{\mathrm{source}\, s} P_s(\psi_i) P_s(\bm r_i | \psi_i)  p_{i, s}.
    \label{eqn:likelihood}
\end{equation}
For event $i$, $P_s(\psi_i)$ and $P_s(\bm r_i | \psi_i)$ are the PDFs of EVPA $\psi_i$ and position on the sky $\bm r_i$, assuming they originated from source $s$, while $p_{i, s}$ is the probability that $i$ was emitted by $s$. We use ``source'' to refer to any potential origin of an IXPE event, which may include astrophysical sources, photon background, or particle background. In Eq.~\ref{eqn:likelihood}, $P_s(\psi_i)$ and $P_s(\bm r_i | \psi_i)$ act as polarization (\S\ref{sec:w-pol}) and spatial weights (\S\ref{sec:w-psf}). $p_{i,s}$ may be used to de-weight particle-like events (\S\ref{sec:w-bg}) and incorporate spectral and temporal weights (\S\ref{sec:w-spec}).

The maximum likelihood estimator (MLE) of parameter $\theta$ is the value that maximizes $L$. With many events, MLE methods are unbiased and deliver optimal uncertainties. Here we estimate the covariance matrix $\Sigma$ by computing $(\Sigma^{-1})_{ij} = -\frac{\partial}{\partial \theta_i} \frac{\partial}{\partial \theta_j} \ln L$ at the maximum. The Bayesian approach with wide, uniform priors in the source Stokes coefficients is equivalent if Gaussian posteriors are assumed, though this assumption may be lifted using a Markov-chain Monte Carlo (MCMC) method. Tools for both types of fits are provided as updates to the \texttt{LeakageLib} package, originally designed to treat polarization leakage \citep{dinsmore2024leakagelib}.\footnote{\url{https://github.com/jtdinsmore/leakagelib}}

\subsection{Polarization Weights}
\label{sec:w-pol}
We define $q_i=\cos 2\psi_i$ and $u_i = \sin 2\psi_i$ as the Stokes coefficients of event $i$.\footnote{Our definition of $q_i$ and $u_i$ differs from that of the IXPE level 2 files, which report Stokes coefficients as $2\cos 2\psi_i$ and $2\sin 2\psi_i$.} The PDF of $\psi_i$ is commonly modeled as
\begin{equation}
    P_s(\psi_i) = \frac{1}{2\pi}\brackets{1 + \mu_i\parens{Q_s q_i + U_s u_i}},
    \label{eqn:pol-distro}
\end{equation}
where $Q_s=\Pi_s \cos 2\Psi_s$ and $U_s=\Pi_s \sin 2\Psi_s$ are the source's Stokes coefficients, and $\Pi_s$ and $\Psi_s$ are its polarization degree (PD) and angle (PA).

The modulation factor $\mu_i$ is the PD detected from similar quality events emitted by a 100\% polarized source. Such $\mu_i$ are provided by NN reconstruction \citep{peirson2021deep}, using the full event morphology to assess the quality of the polarization measurement. Modulation factors are often approximated by the mean response to events of similar energy $\mu_i = \mu_{E_i}$, as measured from simulation or calibration data, but these lack the event quality assessment.

A common analysis technique is to fit a source polarization to all events within an aperture, otherwise disregarding spatial information by considering only this $P_s(\psi_i)$ term. In the low PD limit of this case, the polarization MLE is analytic:
\begin{equation}
    \hat Q_\mathrm{src} = 2\frac{\sum_i \mu_i q_i}{\sum_i \mu_i^2},\qquad \hat U_\mathrm{src} = 2\frac{\sum_i \mu_i u_i}{\sum_i \mu_i^2}.
    \label{eqn:weight-no-bg}
\end{equation}
This result has also been found by \cite{marshall2021multiband,bambi2024handbook} for $\mu_i = \mu_{E_i}$, and by the standard \cite{kislat2015analyzing} analysis in the special case of constant $\mu_{E_i}$. Full treatment of the additional terms in Eq.~\ref{eqn:likelihood} offer improvements over these methods.

\subsection{Spatial Weights}
\label{sec:w-psf}
An event arriving from sky coordinate $\bm x$ is offset by the point-spread function (PSF) of the telescope and errors in the track reconstruction method. The PDF of the detected position $\bm r_i$ is thus
\begin{equation}
    P_s(\bm r_i | \psi_i) \propto \int d^2 \bm x\, d^2\bm \delta\, I_s(\bm x)P_\mathrm{mir}(\bm r_i - \bm \delta - \bm x) P(\bm \delta | \psi_i) 
    \label{eqn:prpsi}
\end{equation}
where $I_s(\bm x)$ is the (usually known) unpolarized surface brightness map of the source, $P_\mathrm{mir}$ is the mirror PSF, and $\bm \delta$ is the reconstruction error vector. The reconstruction error PDF $P(\bm \delta | \psi_i)$ depends on the event EVPA, leading to polarization leakage. As in \cite{bucciantini2023polarization}, we Taylor expand $P_\mathrm{mir}$ in $\bm \delta$ since $\delta \ll$ the PSF width. The $\bm \delta$ integral in Eq.~\ref{eqn:prpsi} reduces to the first few moments of $P(\bm \delta | \psi_i)$, whose values and energy dependence were measured in \cite{dinsmore2024polarization}. In particular,
\begin{equation}
    \begin{aligned}
        P_s(\bm r_i | \psi_i) &=C_{i,s}^{-1}\int d^2\bm x\, I_s(\bm x) \bigg[I_0(\bm r_i-\bm x)\\
        &+ Q_0(\bm r_i-\bm x)q_i + U_0(\bm r_i-\bm x)u_i \\
        &+ (q_i^2-u_i^2) X_0(\bm r_i-\bm x) + 2q_i u_i Y_0(\bm r_i-\bm x)\bigg]
    \end{aligned}
    \label{eqn:leakage}
\end{equation}
where $I_0$, $Q_0$, $U_0$, $X_0$, and $Y_0$ are the known functions of position and event energy reported in \cite{dinsmore2024polarization}, and
\begin{equation}
    \begin{aligned}
        C_{i,s} &= \int_A d^2\bm r \int d^2 \bm x\, I_s(\bm x) \bigg[I_0(\bm r - \bm x) \\
        &+ \frac{\mu_i}{2} \parens{Q_s Q_0(\bm r - \bm x) + U_s U_0(\bm r - \bm x)}\bigg]
    \end{aligned}
    \label{eqn:normalization}
\end{equation}
is the normalization constant, with $A$ representing the extraction aperture. Eq.~\ref{eqn:normalization} is obtained by setting $\int_0^{2\pi} d\psi \int_A d^2\bm r\, P_s(\bm r | \psi) P_s(\psi) = 1$.

Spatial weights, represented by the $P_s(\bm r_i | \psi_i)$ term of Eq.~\ref{eqn:likelihood}, are important for observations of faint sources because they help to de-weight background events. They also correct for polarization leakage biases, as we now prove. Consider polarimetry extraction of a bright point source with a small or irregular aperture, where polarization leakage pollutes the spatially unweighted estimator with additional Stokes coefficients $Q_\ell$ and $U_\ell$. The methods of \cite{dinsmore2024polarization} predict
\begin{equation}
    Q_\ell = \frac{\sum_i \mu_i \int_A d^2 \bm r\, Q_0(\bm r)}{\sum_i \mu_i^2},
    \ 
    U_\ell = \frac{\sum_i \mu_i \int_A d^2 \bm r\, U_0(\bm r)}{\sum_i \mu_i^2}.
\end{equation}
for a weakly polarized point source.

Since the source is bright, the data may be modeled by a single component $s=\mathrm{src}$ representing the point source, giving $L = \prod_i P_\mathrm{src}(\psi_i) P_\mathrm{src}(\bm r_i | \psi_i)$. Furthermore, $P_\mathrm{src}(\bm r_i | \psi_i)$ is independent of the fit parameters $Q_\mathrm{src}$ and $U_\mathrm{src}$ except through $C_{i,\mathrm{src}}^{-1}$, and we may drop the constant terms because they do not affect the MLE. Hence $L \propto \prod_i P_\mathrm{src}(\psi_i) C_{i,\mathrm{src}}^{-1}$. For a point source, Eq.~\ref{eqn:normalization} gives $C_{i,\mathrm{src}} = 1 + \frac{\mu_i}{2}\int_A d^2 \bm r\, \brackets{Q_\mathrm{src} Q_0(\bm r) + U_\mathrm{src} U_0(\bm r)}$. Maximizing this simplified $L$ in the low PD limit gives the MLE:
\begin{equation}
    \hat Q_\mathrm{src} = 2\frac{\sum \mu_i q_i}{\sum \mu_i^2} - Q_\ell,
    \qquad
    \hat U_\mathrm{src} = 2\frac{\sum \mu_i u_i}{\sum \mu_i^2} - U_\ell.
    \label{eqn:weight-bg}
\end{equation}

This demonstrates that our MLE technique correctly subtracts polarization leakage effects. This will extend to more complicated scenarios, provided good statistics and valid source flux models, given the unbiased source parameter retrieval of MLE methods.

\subsection{Background Weights}
\label{sec:w-bg}
IXPE data include particle background events, which are often suppressed using cuts outlined in e.g.~\cite{dimarco2023handling}. These cuts are rather weak to minimize loss of true source photons; de-weighting the background can better separate photon and particle signals. We have developed a convolutional neural network (CNN) to provide particle weights. The 28,609 parameter design, built with \texttt{tensorflow}, consists of two convolutional layers, max pooling, two more convolutional layers and a dense layer, all with ReLU activation. A final dense layer with sigmoid activation yields a one-dimensional output from 0 to 1 for each event, which we refer to as the ``particle character'' $\pi_i$. Large $\pi_i$ corresponds to likely particles.

The CNN is trained with binary cross-entropy loss on data sets extracted from IXPE observations of bright, low-polarization point sources (observation IDs 01002401, 01002601, 02002399, and 02001901). The source $S$ data set consists of events within $26''$ of the source, while the equal-size background $B$ data set contains events beyond $117''$. We take equal sampling of the 1-10 keV energy range in both data sets so that the trained CNN will not exhibit energy-dependent bias. But the background $B$ is inevitably polluted by wings of the central point source and astrophysical photon backgrounds. Training on $B$ and $S$ gives an initial CNN, used to create a nearly photon-free $B'$ by removing low $\pi_i$ (photon) events from $B$. We then retrain on $S$ and $B'$. The re-trained CNN performs quite well; $95\%$ of validation events in the $B'$ (photon-cleaned) data set are cut with a $\pi_i < 0.5$ threshold. Only 5\% of events were misclassified, assuming $B'$ is perfectly clean. This final CNN and a script to tag IXPE level 2 files with $\pi_i$ is provided in the updated version of \texttt{LeakageLib}. 

\begin{figure*}
    \centering
    \includegraphics[width=\linewidth]{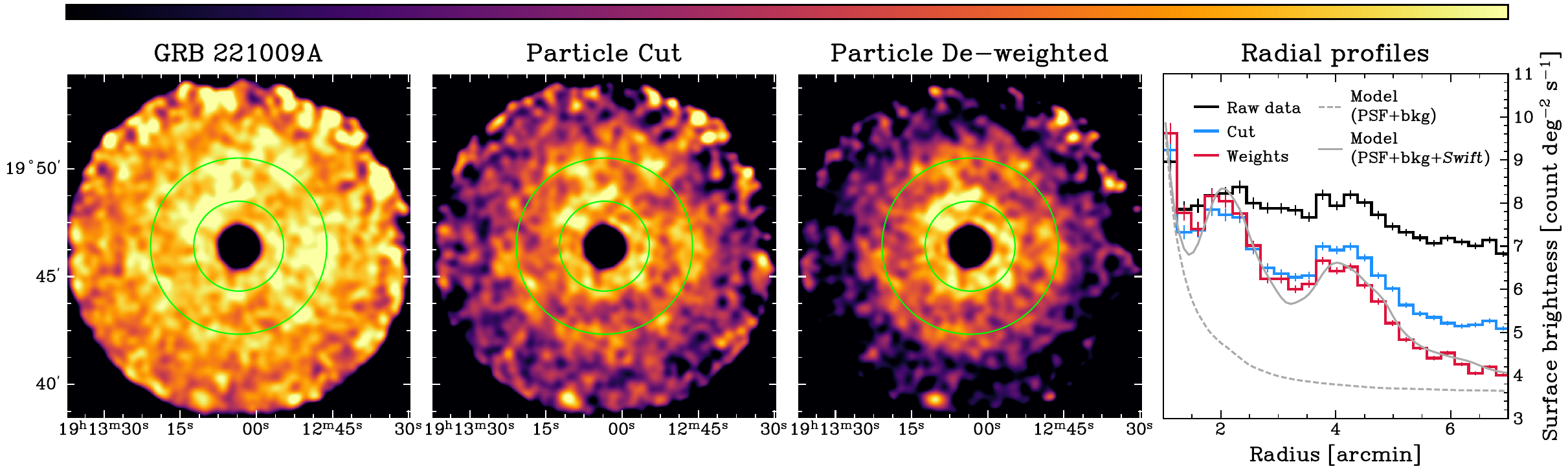}
    \caption{The halo representing scattered prompt emission from GRB 221009A with different background removal techniques. \textit{From left}: all $2-8$ keV counts; counts surviving the background cut of \cite{dimarco2023handling}; counts weighted by the photon character $1-\pi_i$ this work introduces; radial profiles of the images. Green circles highlight the rings as a visual aid. The stretch, max, and min are identical between all panels. The gray curve in the rightmost panel presents a model of the true radial profile derived from the \textit{Swift} GRB 221009A observation, accounting for the IXPE PSF and adding a uniform background.}
    \label{fig:grb}
\end{figure*}

\begin{figure}
    \centering
    \includegraphics[width=0.9\linewidth]{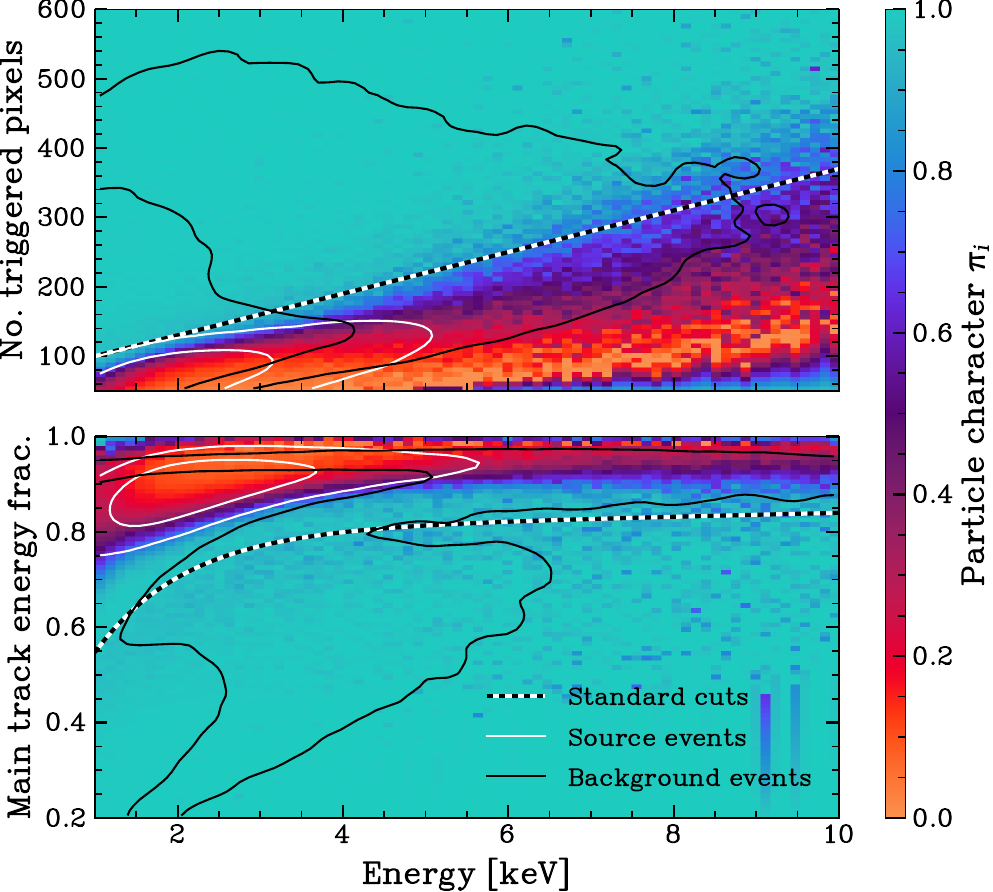}
    \caption{The average CNN particle character plotted vs.~Mom track properties for PSR B0540$-$69. The  dotted black and white line represents the mission-standard background cuts. 68\% and 95\% contours of suspected source and background events are also shown. The CNN identifies similar energy dependence of the particle track properties, but enables the use of more precise weighting methods.}
    \label{fig:dimarco}
\end{figure}

Fig.~\ref{fig:grb} demonstrates the efficacy of particle weights for faint sources using the brightest of all time (B.O.A.T.) Gamma-ray burst as an example. GRB 221009A triggered an IXPE target of opportunity observation \citep{negro2023ixpe} that began just two days after the burst's detection and continued for two more. Echos of the initial explosion were detected as faint rings surrounding the point-like afterglow, reflected by Galactic dust. These rings are much more prominent when particles are de-weighted rather than cut, which will be critical for future polarization analyses. Good agreement between the model GRB radial profile (see appendix \ref{app:grb}) and the IXPE radial profile further illustrates this.

Our CNN assigns high $\pi_i$ to events cut by \cite{dimarco2023handling}. Fig.~\ref{fig:dimarco} shows our particle character estimates in the space of two level 1 parameters used to determine these cuts. Contours show the event distribution within source ($r<10''$) and background regions ($180'' <r<280''$) for the IXPE observations of PSR B0540$-$69. Low $\pi_i$ events are removed from the background contour to clean it of solar photons. Clearly, the additional information of the CNN allows improved background suppression, since some uncut events are flagged with high $\pi_i$. De-weighting these events may incur slight source flux loss; our treatment takes this into account when evaluating PD.

The particle character metric $\pi_i$ output by the CNN is different from the probability that a given event represents a particle $P_{i,\mathrm{ptcl}}$; the latter should take into account the observation's particle-to-photon flux ratio. By Bayes' theorem,
\begin{equation}
    \frac{\pi_i}{1 - \pi_i} \frac{F_\mathrm{ptcl}}{1 - F_\mathrm{ptcl}} =  \frac{P_{i,\mathrm{ptcl}}}{P_{i,\mathrm{phot}}}
    \label{eqn:bayes}
\end{equation}
where $F_\mathrm{ptcl}$ is the fraction of flux stemming from particles. Note that $\pi_i = P_{i,\mathrm{ptcl}}$ only for the training data set's special case of $F_\mathrm{ptcl} = 1/2$.

In maximum likelihood fits, one assigns all sources $s$ an emission probability $p_{i,s} = F_s P_{i,\mathrm{phot}}$ for photon sources and $p_{i,s} = F_sP_{i,\mathrm{ptcl}}$ for a particle sources, where $F_s$ is the source flux fraction. Eq.~\ref{eqn:bayes} allows $P_{i,\mathrm{ptcl}}$ to be given in terms of $P_{i,\mathrm{phot}}$, which may then be dropped as a constant normalizing factor. The final emission probabilities are
\begin{equation}
    p_{i,s} = \begin{cases}
        F_s & \mathrm{for\ photon\ sources} \\
        \frac{\pi_i}{1 - \pi_i} F_s & \mathrm{for\ particle\ sources}\\
    \end{cases}.
    \label{eqn:pis}
\end{equation}
We fit for $F_s$ under the $\sum_s F_s = 1$ constraint.

\subsection{Spectral and Temporal Weights}
\label{sec:w-spec}
MLE methods can suffer from bias when applied to observations of multiple bright sources with distinct spectra. In such cases, the events from each source exhibit different modulation factor distributions and the correct distribution must be used to extract polarizations for a given source.
For example, consider a polarization MLE for a point source embedded in an unpolarized background. A naive approach would estimate aperture polarization using Eq.~\ref{eqn:weight-no-bg}, measure the counts $N_\mathrm{src}$ ($N_\mathrm{bg}$) in source (background) apertures, and multiply the estimator by the background rate $f_N = N_\mathrm{src} / (N_\mathrm{src} - \phi N_\mathrm{bg})$ where $\phi$ is the source-to-background area ratio. While suitable for \texttt{PCUBE}, for MLE methods $f_N$ is incorrect because Eq.~\ref{eqn:weight-no-bg} divides by $\mu^2$ summed over the entire aperture, which includes both source and spectrum-polluting background events. An unbiased approach is to multiply by a modified background rate
\begin{equation}
    f_\mu = \frac{\sum_{i\in \mathrm{src}} \mu_i^2 }{\sum_{i\in \mathrm{src}} \mu_i^2 - \phi\sum_{i\in \mathrm{bg}} \mu_i^2},
    \label{eqn:f-bg}
\end{equation}
where $i\in \mathrm{src}$ and $i\in \mathrm{bg}$ denote events in the source and background apertures. This normalization accounts for the contaminating background $\mu$ population.

For MLE analyses using the full likelihood instead of Eq.~\ref{eqn:weight-no-bg}, we account for spectral differences by replacing the flux $F_s$ in Eq.~\ref{eqn:pis} with a spectrum $F_s(E_i)$, multiplying by instrument response functions to account for effective area and energy resolution. The spectral parameters can in principle be simultaneously fit with the polarization to achieve spectropolarimetric fits---potentially useful for detailed spectral studies of brighter sources. But spectral fitting requires modeling of IXPE's finite energy resolution and is beyond the scope of this work, which focuses on broadband polarimetry of faint sources.

\cite{wong2023improved} explored phase weighting with IXPE in a ``simultaneous fitting'' analysis of the Crab pulsar and PWN. In the maximum likelihood framework, temporal or phase weights may be included by replacing $F_s$ with the light curve $F_s(t_i)$. Phase weighting with this method makes no immediate improvement over \cite{wong2023improved}, but it allows the simultaneous use of other weight techniques which are beneficial. If the expected source polarization is also time-variable, $Q_s$ and $U_s$ may likewise be replaced with the variability model $Q_s(t_i)$ and $U_s(t_i)$. This paper considers only constant $Q_s$ and $U_s$ in a given phase bin, but \texttt{LeakageLib} provides tools to fit for continuously variable source polarizations, as seen in e.g.~IXPE observations of pulsars.

\section{Validation of Spatial and Spectral Weighting on Simulated Data}
\label{sec:injection}
To demonstrate the power of weighted MLE methods, this section compares the Stokes coefficient uncertainties extracted by our analysis and the mission-standard \texttt{PCUBE} algorithm on simulated data. We focus here on spatial and spectral weights, neglecting particle weights since we lack an accurate simulation of particle events.

\texttt{PCUBE} is a component of the \texttt{IXPEobssim} library \citep{baldini2022obssim} and reports polarization 
\begin{equation}
    \hat Q_\mathrm{src} = 2\frac{\sum_i (q_i w_i/\mu_i)}{\sum_i w_i}, \qquad
    \hat U_\mathrm{src} = 2\frac{\sum_i (u_i w_i/\mu_i)}{\sum_i w_i},
    \label{eqn:pcube}
\end{equation}
where $\sum_i$ sums over all $N$ events in a source aperture, and $w_i$ is a per-event weight that could be set to the Mom-estimated track weights in a weighted analysis \citep{dimarco2022weighted}, or unity in an unweighted analysis. We compare with an unweighted analysis in this section, given the difficulty of simulating $w_i$ without a fully detailed \texttt{Geant4} simulation. This is a reasonable step because the next section demonstrates on real data that the use of Moments weights decreases PD errors by at most $\sim 6\%$. To remove pollution from the unpolarized background, we multiply Eq.~\ref{eqn:pcube} by $f_N$ defined in \S\ref{sec:w-spec}. Uncertainties on $\hat Q_\mathrm{src}$ and $\hat U_\mathrm{src}$ are obtained by propagation of errors. Our simulation is of a highly polarized (PD$=50\%$) point source, with details in appendix \ref{app:sim}. The simulated background is unpolarized, but we allow the background polarization to vary in our fit when extracting source polarizations as one would when analyzing a real data set.

\begin{figure}
    \centering
    \includegraphics[width=\linewidth]{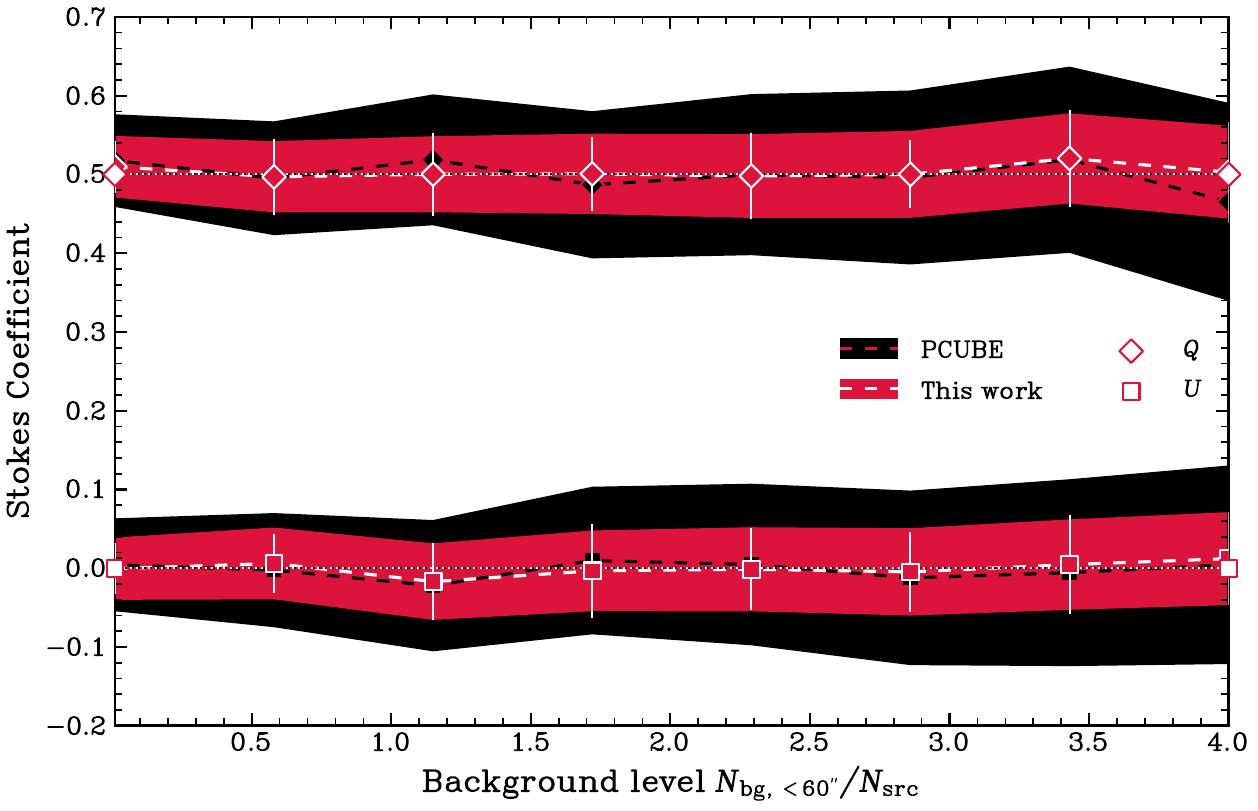}
    \caption{Uncertainties as a function of background flux, measured by the ratio of background counts within the 60$''$ \texttt{PCUBE} radius to source counts. The number of source counts is kept constant. Dotted lines show the true Stokes coefficients, and dashed lines show the coefficients extracted by the two fit methods. Bands reflect the estimated uncertainties. Especially for faint sources, our method delivers substantially more precise fits than \texttt{PCUBE}.}
    \label{fig:bg}
\end{figure}

Fig.~\ref{fig:bg} shows how background flux affects the accuracy of standard \texttt{PCUBE} (black) and our PSF-weighted analysis (red). Dashed lines and bands give the best-fit values and 1$\sigma$ uncertainties, averaged over many simulated data sets. Both methods produce best-fit Stokes coefficients in agreement with the true values. We also measure the standard deviation of best-fit Stokes coefficients from our method (white error bars). This constitutes a measurement of our method's uncertainties; Fig.~\ref{fig:bg} shows they agree with the estimated uncertainty (red band).

For very bright (low background) sources, spatial weighting presents a modest improvement in precision over \texttt{PCUBE}. This improvement is largely due to the bright-source MLE method  (Eq.~\ref{eqn:weight-no-bg}), which outperforms the \texttt{PCUBE} approach (Eq.~\ref{eqn:pcube}). The next section demonstrates, using real IXPE data, that this improvement over \texttt{PCUBE} occurs even when Moments weights $w_i$ are used. As the background increases, \texttt{PCUBE} uncertainties (and uncertainties from any other aperture-based method) grow, even after background subtraction. However, the spatial weighting strongly suppresses this background so that uncertainties increase very slowly. Thus for faint (especially extended) sources, spatial weighting provides a dramatic improvement.

These simulations are ideal cases, using IXPE's on-orbit PSFs measured from bright sources. But the IXPE aim-point wanders by up to tens of arcseconds over an orbit due to boom drift. In standard processing of some sources, this drift is analytically modeled and corrected using the HEASoft \texttt{ixpeboomdriftcorr} tool, but using bright sources we see that the modeling can be imperfect, and we can follow the source motion to remove the residual errors. We measured such residual boom drift patterns for several bright sources and used these to corrupt our simulated data. Even in cases where residual drift reached 17$''$ (a maximal value), PSF fitting gave Stokes coefficients well within 1$\sigma$ of the true values, and uncertainties were only slightly inflated. In dealing with real data, using the bright PSFs should be acceptable, with a Gaussian `blur' added to the PSF if necessary to account for residual boom drift and any small unresolved source extension.

\begin{figure*}
    \centering
    \includegraphics[width=\linewidth]{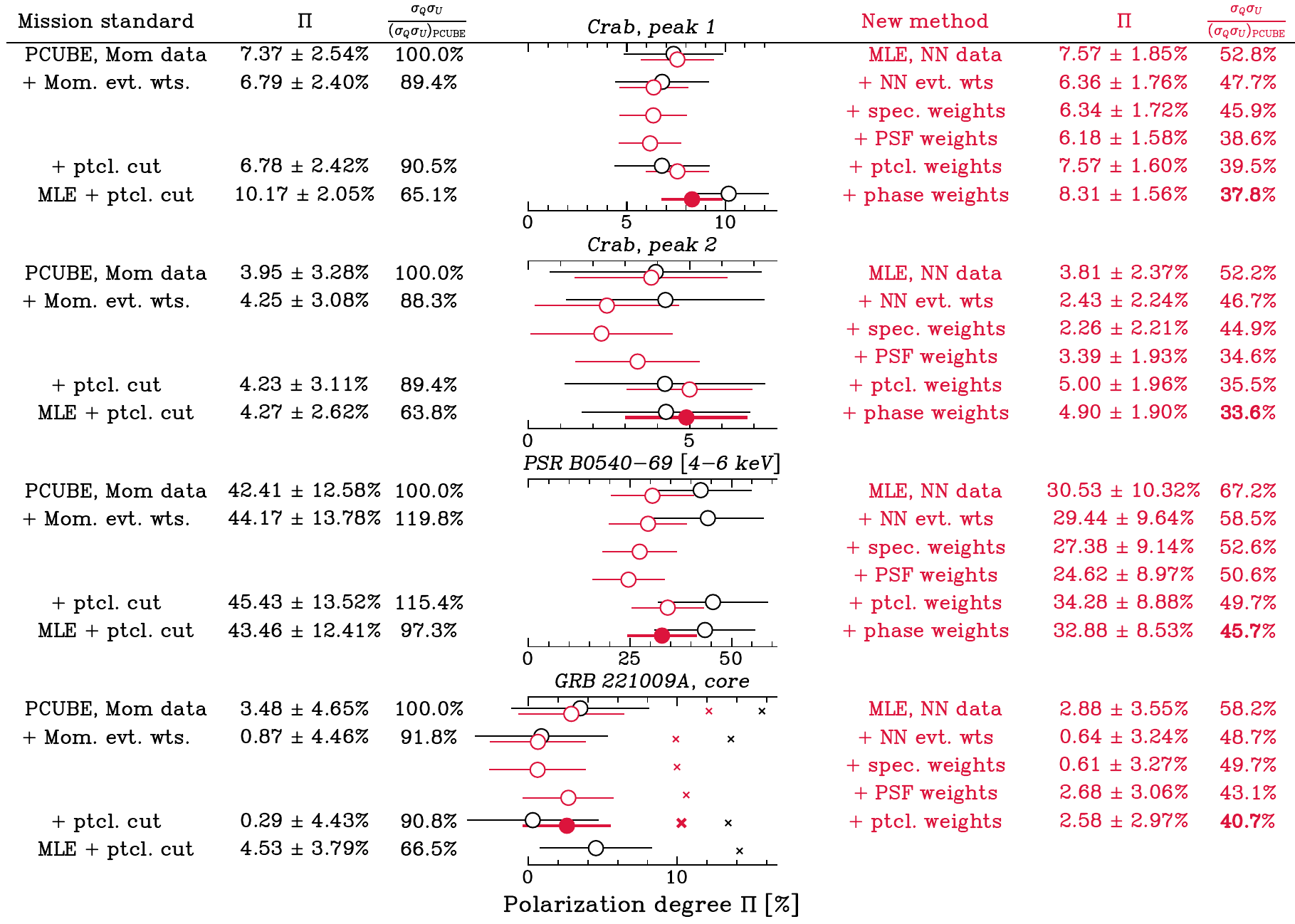}
    \caption{Polarization degrees extracted for four point sources using aperture-based methods in common use (\textit{left}) and the weights considered in this work (\textit{right}). Black results use Mom-reconstructed data while red results use NN reconstruction. Uncertainties are expressed in terms of the PD uncertainty and the area of the Q-U uncertainty contour ($\sigma_Q \sigma_U$) relative to the \texttt{PCUBE} area. PD Results are compared in the middle column. The weights introduced in this paper together substantially reduce polarization uncertainties. For the GRB, $\times$s indicate 99\% confidence upper PD limits since polarization is not detected.}
    \label{fig:table}
\end{figure*}

\section{Application to IXPE Data}
\label{sec:data}

To validate weights performance with real data, we compare the polarization sensitivity of standard analyses with that of our new techniques for three previously analyzed IXPE sources. These are the bright Crab pulsar (PSR B0351$+$21) and its resolved PWN, the ``Crab twin'' (PSR B0540$-$69, hereafter B0540) in the large Magellanic cloud and its unresolved PWN, and GRB 221009A. IXPE observations of these have been well described in earlier publications \citep[respectively]{bucciantini2023simultaneous,xie2024first,negro2023ixpe}. We report uniform polarizations for the central point source over a broad energy range, although each of these objects also includes more complex spatial, spectral, and temporal polarization behavior. We discuss only polarization sensitivity here and do not attempt detailed analysis of the signals. In the conclusions we do comment briefly on the methods' promise for extracting more physics from IXPE data. Descriptions of the analysis for each source are given in appendix \ref{app:fit}.

The comparison is presented in Fig.~\ref{fig:table}. Our figure of merit (FoM) is the area of the polarization ellipse $\sigma_Q\sigma_U$, with decreased area representing increased sensitivity (equivalently, increased effective exposure). It is also important to check that the techniques do not induce undue biases, and so the middle of the figure shows fit PDs and their uncertainty ranges. The left side of the figure shows commonly used aperture-based analysis using Mom reconstruction. On the right we show results from NN reconstructions coupled with new weighting techniques described in this paper.

First on the left, we treat Mom-reconstructed data with unweighted \texttt{PCUBE} (Eq.~\ref{eqn:pcube}, row 1). Polarization sensitivity generally improves when Mom event weights are added (row 2) and when background is suppressed using the cuts of \cite{dimarco2023handling} (row 3). In row 4, we apply the MLE aperture analysis with background rate correction (Eqs.~\ref{eqn:weight-no-bg} and \ref{eqn:f-bg}) instead of \texttt{PCUBE}, keeping the particle cut in place but removing Mom event weights. In all cases, this provides better sensitivity than previous rows. We conclude that the MLE-based analysis using the mean modulation factor is preferred over \texttt{PCUBE} Mom event-weighted analyses. To test whether Mom event weights can improve the MLE analysis, we tried a ``weighted MLE analysis'' (weighting the sums of Eqs.~\ref{eqn:weight-no-bg} and \ref{eqn:f-bg} with $w_i$). The result improves on the \texttt{PCUBE} weighted analysis, but still possesses larger uncertainties than the unweighted MLE by a factor of $10-15$\%.

On the right, we summarize results from the NN-based weighting techniques. First, NN-reconstruction with the MLE method (Eqs.~\ref{eqn:weight-no-bg} and \ref{eqn:f-bg}) and energy-binned modulation factor $\mu_{E_i}$ provides improved sensitivity compared to even the best Mom-bases analysis (row 4, left) and greatly improved sensitivity compared to standard \texttt{PCUBE} (row 1, left). Next, row 2 shows how the sensitivity is further improved by the full event weights $\mu_i$. Note that, unlike Mom event weights, these {\it do} provide an improvement over MLE $\mu_{E_i}$ results, and should be used. The uncertainties for the above analytical MLE methods are obtained from linear propagation of errors.

Row 3 adds spectral weights, which provide some discrimination between sources.\footnote{Astrophysical spectra are obtained from former studies as reported in appendix \ref{app:fit}, the solar photon background is assigned a power law with photon index $\Gamma = 2.5$ which closely matches the data, and NN- and Mom-particle background spectra are measured from data as described in appendix \ref{app:nn}.} This row and those below are obtained by numerically maximizing the likelihood. Row 4 shows how PSF-weighting improves isolation of the point source from polarized background. This PSF analysis also reduces systematic error from leakage effects which occur for small or non-circular apertures. Small-aperture effects are relevant for Crab, but not quantified here. Next, additional improvement is realized by the NN particle background de-weighting. The effect is very small for these point-source extractions---the improvements should be much more substantial for measurement of faint, extended sources. Finally, to illustrate the effect of additional weights, row 5 shows how phase weighting can improve polarization measurements of the pulsar sources in wide phase windows.

Overall one sees that the weighting improves our sensitivity FoM by $\gtrsim 2\times$ over basic \texttt{PCUBE}, the most common IXPE analysis method. The best Mom-based analysis, labeled \texttt{MLE + ptcl.~cut}, has been used in several previous IXPE studies. Our analysis yields a $\gtrsim 30\%$ improvement over that combination. NN analysis has been previously presented for simulated data, so \texttt{MLE, NN data + NN evt.~wts} could have been previously used. However, the NN analysis has not yet been applied to real IXPE data in peer-reviewed publications, so we include it as a new weight here.

A few special cases deserve comment. While MLE analysis typically offers a large gain, for B0540 we have only measured over $4-6$ keV to be consistent with the original publication. In this high-energy band, the modulation factor is nearly constant, so the MLE analysis provides little gain over \texttt{PCUBE}. If the detection band can be widened, improvements should be larger. On the other hand, NN reconstruction performs best for these high energy tracks compared to Mom, so NN uncertainties are much reduced. Also, spatial weighting provides good gain when the source/background ratio flux is small. Crab peak 2 is a good example. As noted, background suppression weights offer modest gain for these bright point sources but can be significant for extended faint sources. Our prescription for the background de-weighting appropriately adjusts to the varying circumstances: when background is weak, particle background is only lightly cleaned (e.g.~Crab). When particle background is strong, then the weights have a larger effect (e.g. the GRB).

\section{Discussion and Conclusions}
\label{sec:conclusion}

This paper highlights the power of NN event analysis and MLE methods to extract polarization measurements from IXPE data. It also introduces new weights which offer additional analysis improvements; detailed instrumental PSFs now allow spatial weights and a new NN tool allows improved characterization and de-weighting of likely particle events. Time/phase and energy weighting can also be applied with this formalism. We have validated our method by comparing it analytically to other methods in appropriate limits, recovering true polarizations from simulated data sets, and re-analyzing a varied set of IXPE observations. The results are consistent with, and improve upon, published measurements and show no measurable bias. Compared to the mission standard \texttt{PCUBE} measurement, we reduce the polarization contour area by $> 2\times$ and make significant improvements to even the best Mom-based polarization analysis. The tools needed to reproduce these results and to apply to other IXPE sources are publicly available in the \texttt{LeakageLib} package.

The most effective weighting component depends on source properties. Maximum likelihood estimates are valuable for analyses over wide energy ranges, where the modulation factor varies dramatically. NN track reconstruction always helps, but is most effective when high energy events contribute significantly. Spatial and particle weighting is most useful in the presence of strong backgrounds; spatial weighting has the additional benefit of separating partially resolved sources and removing polarization leakage effects. Phase weighting is important for rapidly varying sources. We caution that these weights are designed for broad band polarization measurements and are not appropriate for spectral analysis. Indeed, bright sources allowing detailed spectral decomposition will not require these weighting techniques and are currently best dealt with other packages, e.g.~\texttt{3ML} or \texttt{xspec}.

This paper applies spatial weights to point sources (albeit with spatially varied backgrounds), but the methods may readily be applied also to extended sources as well. An analysis of a complex field might proceed as follows: an image delivered by the \textit{Chandra} X-ray observatory gives a high-resolution, unpolarized X-ray image, which the astronomer separates into possibly polarized components. The image is rescaled to the IXPE band, and spatially uniform components are added representing solar photons and particles. If necessary for a good fit, these background components should be modeled as polarized. IXPE polarizations for each source are extracted, using spatial, particle, phase, and/or spectral weights to separate the components. Best results are achieved using NN reconstruction and event weights.

In many cases, IXPE exposures are long and it behooves us to make the best possible use of the collected events. Thus we recommend the use of weighted analysis for all broad-band polarization measurements. Their application represents an increase of $2\times$ or more effective exposure time compared to common unweighted \texttt{PCUBE} measurements. In some cases, e.g.~transients such as GRB 221009A, additional data can never be collected in any case, and weighted analysis can enable additional science. For example, with no significant PD bias, the lowered bounds on afterglow polarization (crosses in Figure \ref{fig:table}) can be used to exclude otherwise acceptable models. Indeed the X-ray PD bounds are now similar to the optical upper limits reported in \cite{negro2023ixpe}. As IXPE attempts to measure ever fainter sources, we expect that the PSF and background weights described here should become increasingly essential.

\begin{acknowledgments}
This work was supported in part by contract NNM17AA26C from the MSFC to Stanford in support of the IXPE project. The authors also thank Niccol\`o Di Lalla, Luca Baldini, and the rest of the IXPE Collaboration for helpful conversations.
\end{acknowledgments}

During the review process, the authors became aware of \cite{ravi2025whats}, which presents a Bayesian analysis tool (\texttt{QUEEN-BEE}) for polarization analysis of a time-varying source. In the special case of a bright point source, the two methods use the same likelihoods and priors and should give identical results. Both methods present likelihood-based analyses as the future of IXPE polarimetry.

\vspace{5mm}
\facilities{IXPE, CXO}
\software{LeakageLib \citep{dinsmore2024leakagelib}, IXPEobssim \citep{baldini2022obssim}, HEASoft \citep{nasa2014HEAsoft}}

\appendix

\section{GRB 221009A Radial Profile Model}
\label{app:grb}
The Neil Gehrels \textit{Swift} observatory measured GRB 221009A's expanding rings of prompt emission throughout IXPE's observation, as \cite{williams2023discovery} reports. The ring reflected by a certain dust cloud expands such that its angular size in radians is $\sqrt {2c\Delta t/D}$, where $c$ is the speed of light, $\Delta t$ is the time since the burst, and $D$ is the distance to the dust cloud  \citep[Eq.~1 of][]{williams2023discovery}. Its intensity is proportional to the Galactic dust density, reduced by photoelectric absorption for rings reflected by more distant dust \citep[Eq.~E1 of][]{williams2023discovery}. This allowed the determination of a line-of-sight dust density profile from \textit{Swift} data. We use this density distribution to predict the IXPE radial intensity profile, integrating the result over IXPE's observing window.

\section{Polarization Leakage Forward Model}
\label{app:sim}
To validate our spatial weights on simulated data in \S\ref{sec:injection}, we construct a simulation capable of modeling the relevant effects, including the telescope response functions, PSF, and polarization leakage. While we cannot use the \texttt{IXPEobssim} simulator because it does not model polarization leakage, we otherwise mirror its approach. We draw event energies from power-law spectra with photon indices of $\Gamma=1.5$ for the source and 2 for the background. The source spectrum is multiplied by the IXPE effective area and quantum efficiency as given in \texttt{IXPEobssim}. The EVPA $\psi_i$ of each event is drawn from the standard Eq.~\ref{eqn:pol-distro} using the modulation factor $\mu_{E_i}$ given in \texttt{IXPEobssim}.

Spatially, background events are distributed uniformly and point source events are initially distributed by the mirror PSFs for each IXPE detector. We use the sky-calibrated PSFs, not the circularly symmetric approximations. The event positions are then offset by a reconstruction error $\bm \delta$ linked to the event EVPA. To determine $\bm \delta$, we use the second moments $\sigma_{\parallel,\perp}^2(E)$ and the fourth moment parallel component $k_\parallel^4(E)$ of the PDF of $\bm \delta$ as determined from on-sky data in \cite{dinsmore2024polarization}. Assuming that all the other moments are Gaussian determines a unique closed functional form for $P(\bm r_i | \psi_i)$: a Gaussian times a series of Hermite polynomials. Values of $\bm \delta$ are drawn from this distribution.

\section{Detailed Analysis Methodology}
\label{app:fit}
This section describes the data reduction and analysis methods used to extract the polarization degrees listed in Fig.~\ref{fig:table}.

\paragraph{Crab} The Crab is bright enough that a single observation (ID 02006001) supplies plenty of signal. We apply the NN-reconstruction method as described in appendix \ref{app:nn} and phase the data after barycentric correction using the CRABTIME ephemeris database. After aligning the detectors to center the pulsar, our polarization is extracted from a 25$''$ aperture during both the ``on'' and ``off'' phases and subtracted to reveal the pulsar polarization. For the spatially weighted methods, we model the nebula within this aperture with a flux map taken from a contemporaneous \textit{Chandra} observation (observation ID 23539), reprocessed, cut to $2-8$ keV, and mapped to the IXPE effective area. For energy weights, our pulsar spectral model is taken from \citep{weisskopf2004chandra} and the nebula spectrum is assumed to exhibit $\Gamma=1.9$, consistent with torus spectra measured in \citep{mori2004spatial}. For phase weights, we measure the light curve from the region of interest and ascribe all the pulsed flux to the pulsar. Since the polarization maps reported in \citep{wong2023improved} found little PA variation within our region, assuming uniform polarization for this structure is acceptable. Spatial variation of the nebular background will produce unphysical polarization leakage signals that will not average to zero inside the aperture. The PSF-weighted analysis automatically corrects for this leakage as demonstrated with Eq.~\ref{eqn:weight-bg}, but the aperture extraction methods do not.

Our measured background (nebula) PA agrees with that previously determined for the center of the Crab \citep{wong2023improved}. For the pulsar, we extract separate PDs for the first (phase $0.17-0.24$) and second ($0.565-0.65$) peaks. Note that more detailed phase analysis shows rapid PA sweeps \citep{bucciantini2023polarization,wong2023improved}, so values are sensitive to the precise boundaries. However our results in these broad bins are consistent with previous measurements and more importantly use identical events for each analysis, allowing useful FoM comparisons.

\paragraph{PSR B0540$-$69} Our B0540 analysis considers IXPE 4-6 keV events from observations 02001201, 02001202, and 02008801, as the original analysis finds significant polarization only in this bin \citep{xie2024first}. When we reprocess NN events, we retain the Mom energy estimates to ensure that we fit to the same events regardless of the reconstruction method.\footnote{This 2\,keV range is not much larger than the IXPE energy resolution, so different event reconstruction can significantly alter the included event set.} We use the same ephemeris, spectrum, and on-off phase bins as that analysis. Our aperture-based methods are conducted with a radius of 60$''$, and the off-phase is taken as background for the on-phase, similarly to Crab. The PSF-weighted analyses use a 278$''$-radius aperture, nearly to the chip edge. Using a large aperture is beneficial because the PSF wings provide some useful counts at these large separations, and spatial weighting ensures the extra background events do not harm the fit precision. For phase weight modeling, we measure the observation's light curve and ascribe all the pulsed flux to the pulsar, plus a DC component with amplitude 8\% of the pulse height \citep{xie2024first}. We see a slight, non-significant drop in PD which would be expected if the PA sweeps over the pulsar period.

The B0540 IXPE image is extended compared to the PSF models. We blur the PSF with a Gaussian, finding a standard deviation of $8.8''$ from a fit to the merged IXPE image. The extension is likely due to imperfect boom correction, with possible contribution from the $\sim 5''$-radius PWN. We treat both the PWN and pulsed emission as point sources under the blurred PSF model, and add faint uniform photon and particle backgrounds. Since a fit shows the backgrounds as consistent with unpolarized to 1$\sigma$, we assume an unpolarized background in the final analysis.

\paragraph{GRB 221009A} Observation 02250101 of the B.O.A.T. was first analyzed using a 26$''$ aperture surrounding the point source \citep{negro2023ixpe}. We use the same radius for aperture-based methods, and assign spectral weights using the spectrum found in that work. Lacking a clean background region, we approximate background flux using a $160''-200''$-radius annulus, which will be slightly polluted by flux from the scattered rings (see Fig.~\ref{fig:grb}). We do not require a clean background region for spatially weighted methods, so for these we remove the background annulus and extend the source aperture to 60$''$. In principle, this source aperture extension risks including some flux from the extended rings of scattered X-rays from the initial explosion. However, at the time of the IXPE observation, the dust responsible for ring emission within our aperture is at distances $>48$ kpc from Earth, where densities are likely quite low (see appendix \ref{app:grb}). Indeed, a fit that includes a ring component shows that its flux within our aperture is consistent with zero. We therefore ignore the rings in the final spatial weighted fit. We also fit uniform background photon and particle components. As with B0540, we see no signal of background polarization and therefore assume that they are unpolarized.

\section{Neural Net Event Reconstruction}
\label{app:nn}
The NN developed and trained in \cite{peirson2021deep} accurately determines the PA, position, and energy of IXPE events based on the ``level 1'' raw track files available in the HEASARC data archive. After processing tracks with the \texttt{ixpeevtrecon} software, we determine NN track properties and run the results through the standard level 1 to level 2 pipeline, shipped through HEAsoft version 6.35.2. Among other effects, this pipeline corrects for telescope gain, boom drift, and spurious modulation. In particular, the spurious modulation effect---a detector artifact yielding spatially dependent anomalous polarization---will differ between Mom and NN reconstructions. Here we use the modulation amplitude scaling in \cite{peirson2021deep} estimated from basic calibration measurements. A more complete analysis of polarized calibration sources in \cite{cibrario2025mitigating} gives similar results.

While the typical difference between NN- and Mom-estimated energies for photon-like tracks is well under 1 keV, on particle-like tracks the methods assign very different energies. From B0540 events with $\pi_i > 0.5$ at distances of $180-280''$ from the source, we measure that the Mom particle spectrum is approximately a $\Gamma = 1.87$ power-law, unmodified by the effective area. The NN particle spectrum is harder and non-power-law. These measured spectra are used for the spectral weighted fits reported in the main text.

\bibliography{bib}{}
\bibliographystyle{aasjournal}

\end{document}